\documentclass[twocolumn]{revtex4}
\usepackage{graphicx}

\begin{document}

\title{Nonlocality of a single particle}
\author{Jacob Dunningham} 
\author{Vlatko Vedral}
\affiliation{School of Physics and Astronomy, University of Leeds, Leeds LS2 9JT, United Kingdom}

\begin{abstract}
There has been a great deal of debate surrounding the issue of whether it is possible for a single photon to exhibit nonlocality. A number of schemes have been proposed that claim to demonstrate this effect, but each has been met with significant opposition.  The objections hinge largely on the fact that these schemes use unobservable initial states and so, it is claimed, they do not represent experiments that could actually be performed. Here we show how it is possible to overcome these objections by presenting an experimentally feasible scheme that uses realistic initial states. Furthermore, all the techniques required for photons are equally applicable to atoms. It should, therefore, also be possible to use this scheme to  verify the nonlocality of a single massive particle. \end{abstract}


\maketitle

John Bell identified nonlocality as the key feature discriminating quantum and classical physics \cite{Bell1964}. The violation of Bell's inequalities \cite{Freedman72, Aspect1981, Aspect1982a, Aspect1982b, Tittel1998, Pan2000} forces us to give up either the view that the world is local or the view that the world is real, i.e. independent of observation. Although nonlocality is widely accepted, there has been a long-standing debate about whether it applies to a single photon. A number of schemes have been proposed to test this case, but they have been criticised for either relying on unobservable initial states or for not representing true single particle effects. Here we show how it is possible to overcome these objections with a scheme that uses realistic states and can be applied to single massive particles as well as photons. This conclusively demonstrates that we must not view nonlocality as pertaining to particles themselves, but see it instead as a property of quantum fields whose significance is, therefore, more fundamental than that of particles.

Feynman once famously claimed that superposition is the only mystery in quantum mechanics. Others would add nonlocality to the list. If, however, single particles can exhibit nonlocality, then these two mysteries become one and the same.
This is an important issue since, in quantum field theory, excitations rather than particles are the most fundamental entities. If nonlocality only existed when we had two or more particles, this would present a serious problem, since there would suddenly be something peculiar about two excitations that would not exist when we had only one. More than 60 years ago, Eddington \cite{Eddington1942} pointed out that quasi-particles, such as Cooper pairs of electrons are as much `particles' as are individual electrons. A single electron should, therefore, be able to exhibit non-locality as much as a Cooper pair.

Up until 1991, any discussion of nonlocality always involved two or more particles. Tan, Walls, and Collett (TWC) were the first to claim that a  single photon could also exhibit this effect \cite{Tan91}.  Hardy \cite{Hardy94} modified their scheme to extend the class of local models it ruled out \cite{Santos92, Tan92}.
He did not, however, manage to stem a growing tide of controversy.
Greenberger, Horne, and Zeilinger (GHZ), in particular, argued ``loudly and clearly" against Hardy's scheme \cite{Greenberger95, Greenberger96}.  They said it did not represent a real experiment and was really a multiparticle effect in disguise.
Despite much debate, there is still no clear consensus
on the matter \cite{TerraCunha06, Bjork01}.

What is needed is an experiment that can unambiguously demonstrate the nonlocality of a single particle. Here we propose just such an experimental scheme by modifying Hardy's work to overcome the concerns of GHZ. In particular, we eliminate any unphysical inputs and consider only mixed and number states. 
A fascinating consequence is that this scheme could be used to verify nonlocality for single particles with mass. Hardy felt that ``... nonlocality with single particles of this type could not be observed" \cite{Hardy94}. 
However, all the techniques employed in our scheme are equally applicable to atoms as to photons and so the results should apply to both.


Let us begin by reviewing the Hardy scheme (see Figure~1). A state, $q|0\rangle + r|1\rangle$, and a vacuum state, $|0\rangle$, are incident on the two input ports of a 50:50 beam splitter. 
The two output modes $u_{1}$ and $u_{2}$ are then respectively combined with local oscillators at two other 50:50 beam splitter and detections are made at the four output ports $c_{1,2}$ and $d_{1,2}$.
We have chosen the particular values $q=1/\sqrt{3}$ and $r=\sqrt{2/3}\,e^{i\phi}$ to simplify the analysis and enable us to develop a specific experimental scheme, but this does not reduce the generality of our  arguments.

\begin{figure}[b]
\includegraphics[width=7.5cm]{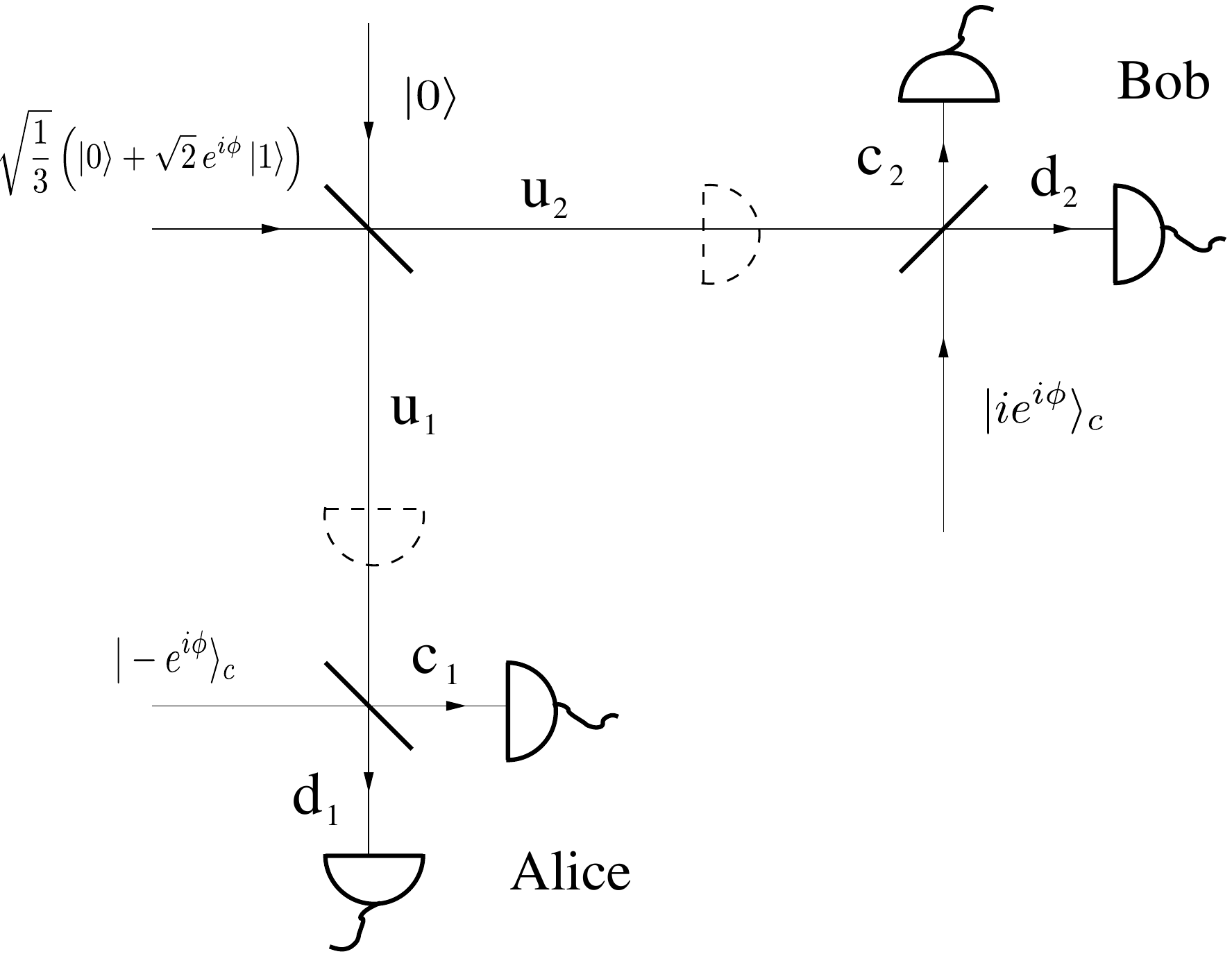}
\caption{The scheme proposed by Hardy for verifying the nonlocality of a single photon.}  \label{chequerfig}
\end{figure}

After the first beam splitter, the state is,
\begin{equation}
|\psi\rangle = \frac{1}{\sqrt{3}}\left[\, |0\rangle |0\rangle +e^{i\phi}\left( |0\rangle |1\rangle + i|1\rangle |0\rangle \right)\right], \label{bs1state}
\end{equation}
where the first ket in each term represents the number of particles on path $u_{1}$ and the second ket represents the number of particles on path $u_{2}$. Now suppose that $u_{1}$ is sent to Alice and path $u_{2}$ is sent to Bob. Alice and Bob each have two choices open to them. They can either directly measure the number of particles on their path  --  represented by the dashed detectors in Figure~1 -- or they can make a homodyne detection by combining their path with a local oscillator at a 50:50 beam splitter -- represented by the solid detectors.
These choices lead to four possible experiments.  

\noindent
{\bf Experiment 1:} Alice and Bob both decide to measure the number of particles on their paths (dashed detectors). In this case, it is clear that they cannot both detect a photon since no more than one photon is emitted from the source at any time. This means that detecting a particle on $u_{1}$ and detecting a particle on $u_{2}$ never happens.

\noindent
{\bf Experiment 2:} Alice elects to make a homodyne detection at  $c_{1}$ and $d_{1}$, by combining the state on path $u_{1}$  with a coherent state, $|-e^{i\phi}\rangle_{c}$ at her 50:50 beam splitter (we will use the subscript $c$ throughout to distinguish coherent states from Fock states).
Bob, meanwhile, makes the same measurement as in experiment 1. If Bob records zero photons on path $u_{2}$, we see
from (\ref{bs1state}) that the state for path $u_{1}$ is,
\begin{equation}
|\psi\rangle = \frac{1}{\sqrt{2}}\left[ |0\rangle +ie^{i\phi}|1\rangle\right]. 
\end{equation}
The output from Alice's beam splitter can then be shown to be,
\begin{equation}
|\psi\rangle =  |0\rangle|0\rangle + i\sqrt{2}\, e^{i\phi}|1\rangle|0\rangle + ... \label{Alice}
\end{equation}
where the first and second kets respectively denote the number of particles on paths $c_1$ and $d_1$ and we have neglected any terms containing more than one particle. We see that if Alice detects a single particle, it must be at $c_1$, since there is no term $|0\rangle|1\rangle$ in (\ref{Alice}). Turning this argument on it's head, if Alice detects a particle at  $d_1$ and nothing at $c_1$, then Bob cannot detect no photons, which means he must detect one (since this is the only other possibility).

\noindent
{\bf Experiment 3:}
The roles of Alice and Bob are reversed: Alice detects the number of photons in path $u_{1}$ and Bob makes a homodyne detection at $c_2$ and $d_2$. Following a similar argument to experiment 2 above, if Bob detects one particle at $d_2$ and nothing at $c_2$, then he can infer that Alice must have detected a particle in path $u_1$.

\noindent
{\bf Experiment 4:} Both Alice and Bob choose to make homodyne detections. One of the possible outcomes of this experiment is that Alice records one particle at $d_1$ and nothing at $c_1$, while Bob records one particle at $d_{2}$ and nothing at $c_{2}$ \cite{Hardy94}.

The result of experiment 4 is rather curious as can be seen by the following argument. Alice infers from her measurement  that a single particle must have travelled along path $u_2$ towards Bob (see  experiment 2). At least, this is true in the sense that, had Bob put his detector in path $u_2$, he would have been guaranteed to detect a particle.
However, at the same time Bob infers from his measurement  that a single particle must have travelled along path $u_1$ towards Alice (see experiment 3). The problem is that they cannot both be right (see experiment 1). So what has led to this contradiction?

Hardy pointed out that this reasoning makes an implicit assumption of locality,  without which there is no contradiction. Alice might deduce from her result that, had Bob measured the number of particles in path $u_{2}$, he would definitely have detected exactly one. However, if Bob had measured $u_{2}$ instead of the homodyne measurement he did make, there might have been a nonlocal influence from Bob's end to Alice's end and then she might have obtained a different measurement outcome.

Though compelling, this argument was met with significant opposition. 
GHZ, in particular, did not like Hardy's introduction of states they termed ``partlycles" -- superpositions of a single particle and the vacuum \cite{Greenberger95, Greenberger96}. They argued that these states were unobservable and could only be part of a real experiment if there were also photons in other modes to ``keep track of the experiment". They proposed a scheme to reproduce Hardy's results that did not need partlycles. However, the additional particles they needed to make the measurement also introduced nonlocality into the system and so it was clear that, in their case, the nonlocality could not be attributed to a single particle.
While it is true that extra particles are required to keep track of the experiment, that does not preclude the possibility of a single particle exhibiting nonlocality. The key is to ensure that the additional particles do not introduce any nonlocality into the system. We now show how this can be achieved.

\begin{figure}[b]
\includegraphics[width=7.5cm]{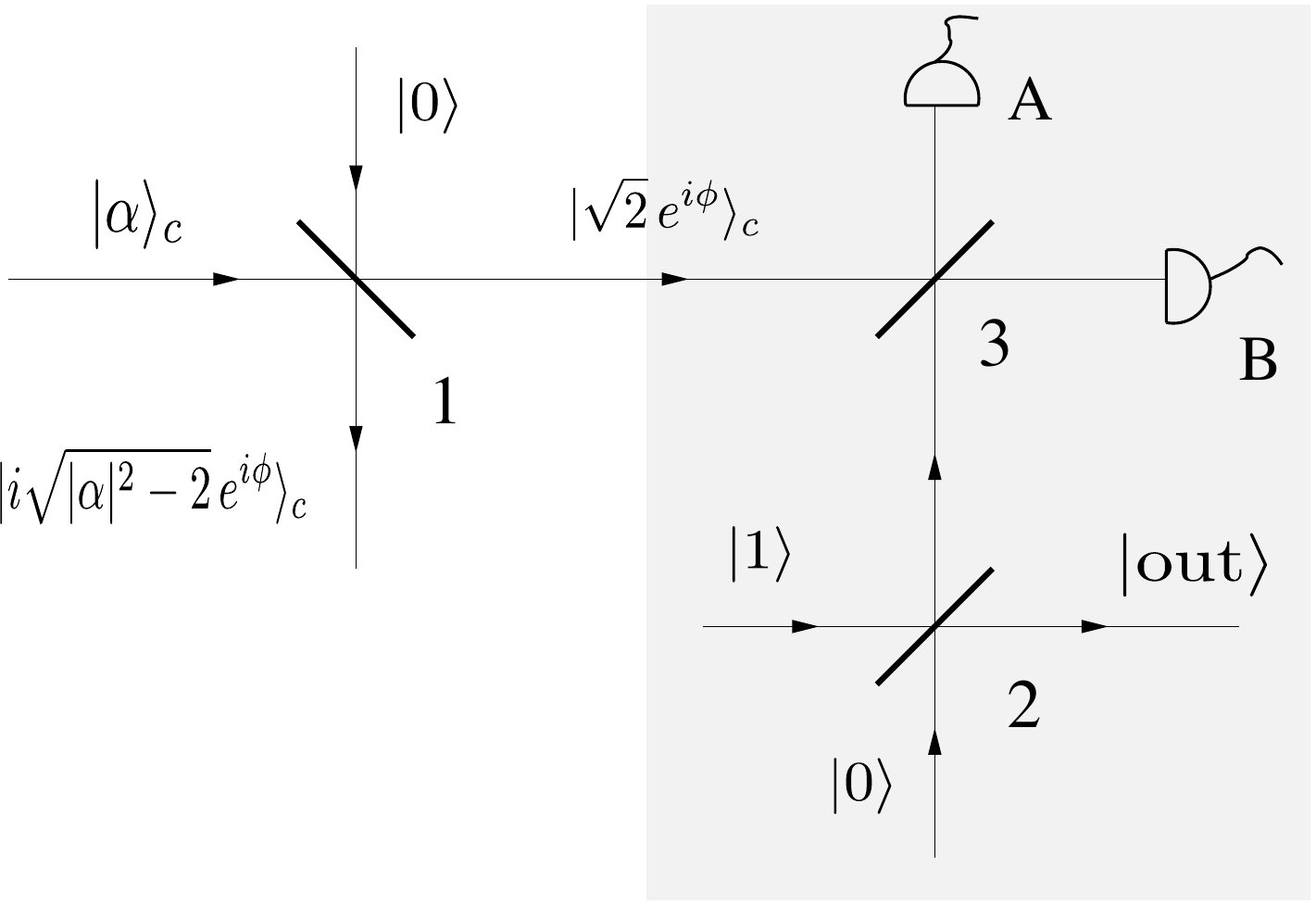}
\caption{(a) Scheme for creating the initial ``partlycle" state. The shaded region is the optical state truncation technique of Pegg {\it et al.} \cite{Pegg98}.  Beam splitter 1 has input states $|\alpha\rangle_{c}$ and $|0\rangle$ and is chosen to have a transmission coefficient of $\sqrt{2}/|\alpha|$. The output state is $|\sqrt{2}\,e^{i\phi}\rangle_{c} |i\sqrt{|\alpha |^{2}-2}\,e^{i\phi}\rangle_{c}$. Beam splitter 2 is a 50:50 beam splitter with inputs $|1\rangle$ and $|0\rangle$.  Outputs from beam splitters 1 and 2 are combined at another 50:50 beam splitter (labelled $3$) and particles are detected at $A$ and $B$. If one particle is detected at $A$ and none are detected at $B$, the output state, $|\rm{out}\rangle$, has the form we want.}  \label{downconv}
\end{figure}

To allay the concerns of GHZ, we will present a modified version of Hardy's scheme that does not rely on states that violate the number conservation superselection rules.  These superselection rules prohibit superpositions, e.g. coherent states or partlycles, but mixed states are always allowed. Our only input states will be the number states, $|0\rangle$ and $|1\rangle$, and the mixed state,
\begin{eqnarray}
\rho  &=& e^{-|\alpha|^2}\,\sum_{n=0}^{\infty} \frac{|\alpha|^{2n}}{n!}\,|n\rangle \langle n|  \label{mixed} \\
 &=&  \frac{1}{2\pi}\int_{0}^{2\pi} ||\alpha|e^{i\phi}\rangle \langle |\alpha| e^{-i\phi}|\, d\phi, \label{mixed2}
\end{eqnarray}
where $\alpha= |\alpha|e^{i\phi}$. 
We see that (\ref{mixed}) can be decomposed into a classical mixture of coherent states averaged over all phases. This means that it is convenient to consider a coherent state with arbitrary phase, $\phi$, and then average over $\phi$ at the end. 
We stress that these coherent states are just calculational tools that will be eliminated at the end of the calculation. We never {\it physically} rely on them.

The input state $\left[ -|0\rangle + \sqrt{2}e^{i\phi}|1\rangle\right]/\sqrt{3}$ \footnote{This apparent partlycle state is simply a consequence of the decomposition. When all the phases are averaged over at the end of the calculation to give the true physical state, this will become a mixed state} can be created by the method of quantum state truncation \cite{Pegg98} as described in Figure~2. 
 We will consider only the case that we detect one particle at $A$ and none at $B$ (the states for all other measurement results will be discarded). In this case, the total state of the system is projected onto \cite{Pegg98},
\begin{equation}
|\psi\rangle = |i\sqrt{|\alpha |^{2}-2}\,e^{i\phi}\rangle_{c}\otimes \left(|0\rangle + \sqrt{2}e^{i\phi}|1\rangle\right).
\end{equation}
The first ket is the `left-over' output from beam splitter 1 and will form our {\it reference} state. The second ket is $|\rm{out}\rangle$ in Figure~2 and has the form we want. Importantly, these two states are not entangled with one another and so they share only classical correlations. 

\begin{figure}[t]
\includegraphics[width=8.5cm]{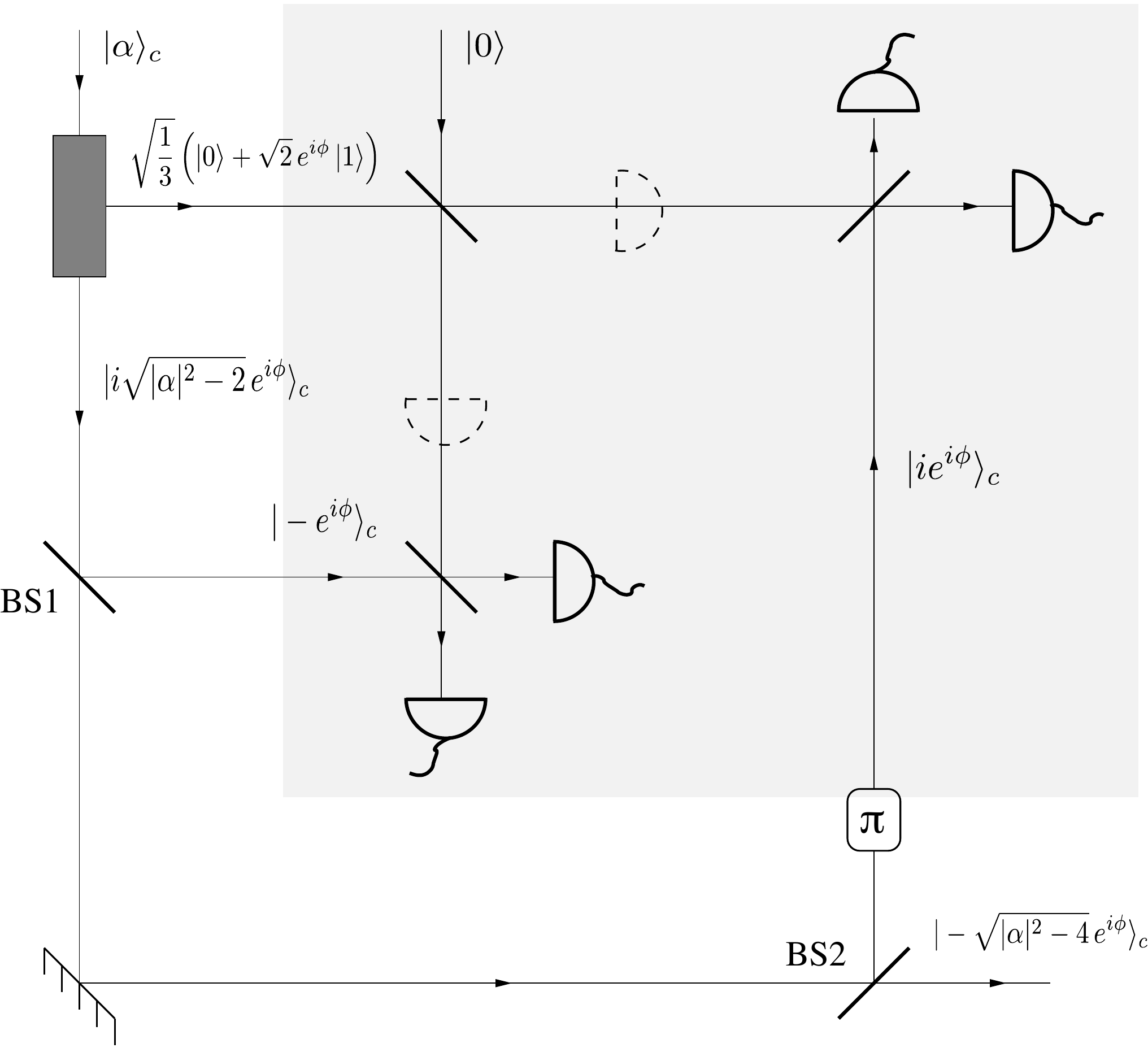}
\caption{Modified Hardy scheme. The shaded area is the original Hardy scheme as depicted in Figure~1. The state creation process depicted by the `black box' is shown in Figure~2. The two beam splitters (BS1 and BS2) are chosen to have reflection coefficients $i/\sqrt{|\alpha|^2 -2}$ and $i/\sqrt{|\alpha|^2-3}$ respectively.}  \label{hardyref}
\end{figure}

The key to our scheme is to use the reference state to provide the local oscillators for Alice and Bob's homodyne detections. 
This  technique of using a common phase reference for both the preparation and measurement of a state is standard  in atom-optics experiments. Here we show that it still works even when we average over all phases. This offers extra justification why, in general, we do not need to introduce additional nonlocality in order to test nonlocality. The full scheme is described in Figure~3. For comparison, the original Hardy scheme from Figure~1 is highlighted in the shaded area.

The first beam splitter that the reference state encounters (BS1) is chosen to have reflection coefficient $i/\sqrt{|\alpha|^2 -2}$, which means it splits off, on average, one particle. The factor of  $i$ is just the $\pi/2$ phase change due to reflection. The output state from BS1 can then be shown to be $|\psi\rangle = |-e^{i\phi}\rangle_{c}|i\sqrt{|\alpha|^{2}-3}\,e^{i\phi}\rangle_{c}$. These outputs are not entangled. The first part, $|-e^{i\phi}\rangle_{c}$, is sent towards Alice and is precisely the state she requires for homodyne detection. The second part is reflected by a mirror to BS2.

BS2 is chosen to have a reflection coefficient of $i/\sqrt{|\alpha|^2-3}$. This again ensures that, on average, one particle is reflected away from the reference state. The output from BS2 can be shown to be $|\psi\rangle = |-ie^{i\phi}\rangle_{c}|-\sqrt{|\alpha|^{2}-4}\,e^{i\phi}\rangle_{c}$ (see Figure~3).  The first part is sent to Bob and, after a phase shift of $\pi $ becomes
$|ie^{i\phi}\rangle_{c}$, which is the state he requires for homodyne detection. The second part, 
$|-\sqrt{|\alpha|^{2}-4}\,e^{i\phi}\rangle_{c}$, is what remains of the reference state.

Overall, the state of the system just before it enters Alice and Bob's homodyne detection beam splitters is,
\begin{eqnarray}
|\psi\rangle = && \left[\, |0\rangle |0\rangle +e^{i\phi}\left( |0\rangle |1\rangle + i|1\rangle |0\rangle \right)\right]\otimes \nonumber \\
&&|-e^{i\phi}\rangle_{c}\otimes |ie^{i\phi}\rangle_{c}\otimes |-\sqrt{|\alpha|^{2}-4}\, e^{i\phi}\rangle_{c}.
\label{overall}
\end{eqnarray}
The first factor, in square brackets is the state of paths $u_1$ and $u_2$ given by Eq.~(\ref{bs1state}). The remaining three factors are respectively the other inputs to Alice and Bob's beam splitters and the remaining output from BS2. These three last states are neither entangled with each other nor with the state in square brackets. They contain only classical correlations and so do not introduce any additional non-locality.
The last step is for Alice and Bob to each perform a local unitary operation (beam splitter) on the state (\ref{overall}). This cannot create nonlocality between them and means that the observed nonlocality can only be due to the state in square brackets, i.e. the single particle state.

Crucially, all of these results are independent of the phase, $\phi$. This means that the results will not change if we average over all phases. In other words, this  scheme should also show evidence of single-particle nonlocality when the only inputs are number  states and mixed states and could, in principle, be carried out in the laboratory. 
When we average over all phases, the `partlycle'  input also becomes a mixed state and so our scheme does not, in any way, rely on these unobservable states.
This means that the objections of GHZ are overcome and the single particle nonlocality  must be taken seriously.

Both Hardy and GHZ concluded (for different reasons) that their schemes could not  verify nonlocality for single massive particles.  By contrast, all the techniques employed in our scheme apply equally well to both atoms and photons: beam splitters with variable reflectivities have been realised for atoms \cite{Keith91,Bongs99, Cassettari00} and spectacular progress has been made in the ability to detect individual atoms. This means that there is no fundamental reason why our scheme could not also be used to verify the nonlocality of a single massive particle.

Finally, we note that one of the pillars of Leibniz's  metaphysics, the principle of identity of indiscernibles, states that objects cannot differ solely in number. This seems to be confirmed by the traditional view of nonlocality where we require two or more particles to violate Bell's inequalities. Here we have shown that this view is incorrect and that even a single particle exhibits nonlocality. This strengthens our belief that the world described by quantum field theory, where fields are fundamental and particles have only a secondary importance, is closer to reality than might be expected from a na\"ive application of quantum mechanical principles.
\newline
{\bf Acknowledgements:}
This work was financially supported by the United Kingdom EPSRC and the Royal Society and Wolfson Foundation.

\end{document}